\documentstyle[prb,aps,epsf,twocolumn]{revtex}

\begin{document}

\title{
Dynamical density--density correlations in one--dimensional Mott insulators
}

\author{
Walter Stephan and Karlo Penc\cite{*}}

\address{Max-Planck-Institut f\"ur Physik komplexer Systeme, Bayreuther
Str. 40, D-01187 Dresden, Germany 
}

\date{September 11,1996}
\maketitle

\begin{abstract}
The dynamical density--density correlation function is calculated for the 
one--dimensional, half--filled Hubbard model extended with nearest 
neighbor repulsion using
the L\'anczos--algorithm for finite size systems and analytically for large 
on site repulsion compared to hopping amplitudes.  
At the zone boundary an excitonic feature exists for any finite
nearest neighbor repulsion and exhausts most of the spectral weight,
even for parameters where no exciton is visible at zero momentum.
\end{abstract}


\narrowtext
Materials such as $SrCuO_2$ and $Sr_2CuO_3$
are believed to be quite well described by the 
one--dimensional half--filled Hubbard model \cite{SrCuO}.
With the current level of
experimental technology it is possible to measure
dielectric response (which is directly related to the density
response function) with good resolution in both momentum
and energy transfer\cite{eels,x-ray}, which makes it of great interest to
understand the response expected from this simplest model
of a Mott--Hubbard insulator.
In this paper we present the combined results of analytic and
numerical calculations which clarify the nature of the dynamical
density response function of the Hubbard model including also
nearest--neighbor repulsion, which is a first step  to
the inclusion of the long--range Coulomb interaction.  
The results presented may serve as a indication of the
appropriateness of the model to particular materials once
experimental data become available.
Previous work of Mori, Fukuyama and Imada\cite{fukuyama} in this 
direction was
based on an effective low--energy model whose relation to our
extended Hubbard model is non-trivial. 

The extended Hubbard model is defined as 
\begin{equation}
 {\cal H} = -t \sum_{j,\sigma} 
  \left( 
    c^{\dagger}_{j+1,\sigma} c^{\phantom{\dagger}}_{j,\sigma} 
    + {\rm h.c.}
  \right)
  + {\cal U} + {\cal V},
\end{equation}
where ${\cal U} = U \sum_{j} n_{j,\uparrow} n_{j,\downarrow}$
is the on--site Coulomb repulsion,
${\cal V} = V \sum_{j} n_{j} n_{j+1}$ is the nearest--neighbor repulsion and 
$n_i=n_{i\uparrow}+n_{i\downarrow}$ is the density at site $i$.

 The imaginary part of the density-density correlation function is given by
\begin{equation}
  {\cal N}(k,\omega) = \sum_f 
     \left| \langle f | n_k | {\rm GS} \rangle \right|^2 
       \delta(\omega - E_f +E_{\rm GS}) ,
   \label{eq:nko}
\end{equation}
where $| {\rm GS} \rangle$ is the half--filled ground state and
$| f \rangle$ denotes a final state with energy $E_f$ and momentum $k=P_f$. 
Note that the optical conductivity is directly
related to the density--density correlation function through the
 continuity equation (see e.g. Ref.~[\onlinecite{nakajima}])
\begin{equation}
{\rm Re} \> \sigma(\omega) = 
  \lim_{k\rightarrow 0} 
  \frac{\omega}{4 \sin^{2}(k/2)} {\cal N}(k,\omega) .
\end{equation} 
For the case of small charge gap compared to bandwidth $\sigma(\omega)$ has 
been calculated by Giamarchi and Millis \cite{giam} using bosonization, and 
the large $U$ case was studied by Gebhardt {\it et al.}\cite{florian}

We evaluate 
Eq.~(\ref{eq:nko}) by: a) direct diagonalization, 
which is limited to small clusters; b) for $U\gg t,V$ we perform a canonical
transformation of the type introduced by Harris and Lange\cite{harris} 
to get a ``$tJ$'' like effective model. This model is then diagonalized 
for larger system sizes; c) finally, for the effective model in the 
thermodynamic limit an analytic expression is obtained using  
the factorized wave function of Ogata and Shiba\cite{bogyo}.   

\begin{figure}
\epsfxsize=8.5 truecm
\centerline{\epsffile{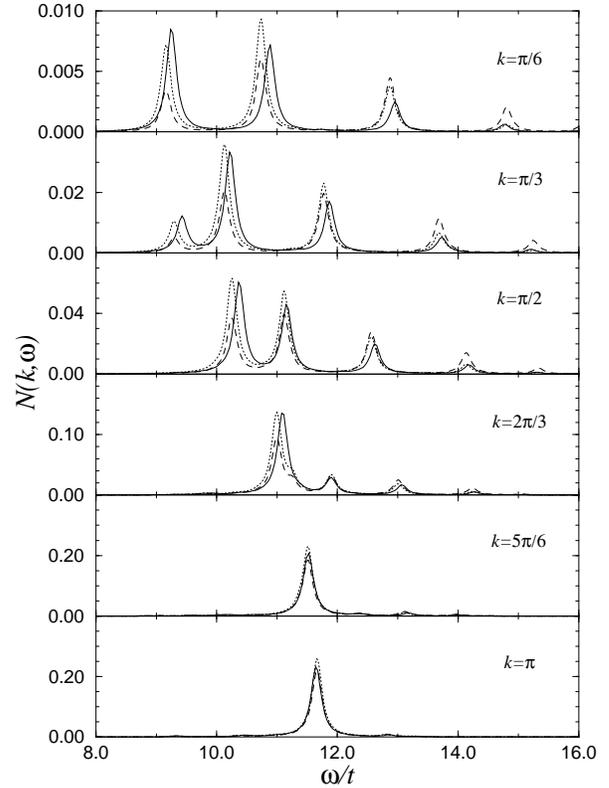}}
\caption{
  ${\cal N}(k,\omega)$ of a half--filled twelve site Hubbard model 
(full line) for $V=t$, $U/t=12$, effective model with 
(dotted line) and without (dashed line) $t^2/U^2$ corections in the density
operators. The $\delta$ functions are plotted as Lorentzians of width $0.1t$.
}
\label{fig:1}
\end{figure}

\paragraph{Exact diagonalization of Hubbard model:}
  Using the standard L\'anczos algorithm, it is straightforward to calculate 
${\cal N}(k,\omega)$ for finite size clusters of $L$ sites. 
In Fig.~\ref{fig:1} we 
show a typical plot for relatively large $U$ and intermediate $V$. 
There are several features to observe:
the spectra consist of dominant peaks which are distributed over a 
frequency range from approximately $U-V-4t$ to $U-V+4t$
 at small momentum transfer, narrowing to a single peak
near $U-V$ at the zone boundary.  There are also less intense
features which are barely visible in the figure whose weight 
increases slightly for smaller $U$.
The number of large peaks at small momentum scales linearly with the 
system size, and is therefore most naturally interpreted as the
finite-size precursor to a continuum absorption.  Near the zone
boundary however, for all accessible system sizes we see only a single
peak when $V$ is finite.  We shall see that this is due to
the appearance of an exciton below the continuum. 
A further feature of ${\cal N}(k,\omega)$ is the asymmetry in the
weight distribution:  the spectra are skewed toward lower frequency,
which is more pronounced for either smaller $U$ or greater $V$.
Our main interest here is the region $V \ll U$, however we
note that for $V$ approaching $U/2$ the transition to the
charge density wave state\cite{bari}  is directly seen as a drastic 
softening of the response at $k=\pi$. 
Apart from the features presented there is of course also
the special case of $k=0$ where the response is simply a
delta function at zero frequency.

\paragraph{Effective model:} In order to understand more clearly the
above observations it is instructive to derive an effective
model in the strong--coupling large $U$ limit.
We closely follow the approach of Harris and Lange\cite{harris}, however, 
for details and notation we refer to Ref. [\onlinecite{oles}].
We first define
\begin{eqnarray}
  \tilde T_{0} &=& \tilde {\cal V} -t \sum_{i,\delta,\sigma} 
      \tilde n_{i,\bar\sigma} 
      \tilde c^{\dagger}_{i,\sigma}
      \tilde c^{\phantom{\dagger}}_{i+\delta,\sigma}
      \tilde n_{i+\delta,\bar\sigma} \nonumber\\
       && -t \sum_{i,\delta,\sigma} 
      (1-\tilde n_{i,\bar\sigma}) 
      \tilde c^{\dagger}_{i,\sigma}
      \tilde c^{\phantom{\dagger}}_{i+\delta,\sigma}
      (1-\tilde n_{i+\delta,\bar\sigma}) \nonumber\\
  T_{U} &=& -t \sum_{i,\delta,\sigma} 
      \tilde n_{i,\bar\sigma} 
      \tilde c^{\dagger}_{i,\sigma}
      \tilde c^{\phantom{\dagger}}_{i+\delta,\sigma}
      (1-\tilde n_{i+\delta,\bar\sigma}),
\end{eqnarray}
and $\tilde T_{-U} = \tilde T_{U}^\dagger $, where the subscript denotes
the change in the eigenvalue of $\tilde {\cal U}$ induced, corresponding
to the change in the number of doubly occupied sites,
so that
\begin{equation}
  {\cal H_{\rm eff}} = \tilde {\cal U} + \tilde T_{0} + \frac{1}{U} 
   \left[\tilde T_U,\tilde T_{-U} \right] 
   + {\cal O}\left( \frac{t^3}{U^2} , \frac{V t^2}{U^2} \right).
\end{equation}
Here by $\tilde O$ we denote the canonically transformed operators.
The expression for ${\cal H_{\rm eff}}$ at the operator level is quite 
long and complicated, here we rather give the effect of
$U^{-1}[\tilde T_U,\tilde T_{-U}] $ 
applied to the basis states:
\begin{eqnarray}
  |\sigma\bar\sigma\rangle &\rightarrow& 
  (2t^2/U) \left(
|\bar\sigma\sigma\rangle - |\sigma\bar\sigma\rangle\right)
  \nonumber\\
  |\tau\bar\tau\rangle &\rightarrow& 
  (2t^2/U) \left(
|\tau\bar\tau\rangle + |\bar\tau\tau \rangle\right)
  \nonumber\\
  |\tau\sigma\bar\sigma\rangle &\rightarrow& 
  (t^2/U) \left(
|\bar\sigma\sigma\tau\rangle - |\sigma\bar\sigma\tau \rangle\right)
  \nonumber\\
  |\sigma\bar\sigma\tau\rangle &\rightarrow& 
  (t^2/U) \left(
|\tau\bar\sigma\sigma\rangle - |\tau\sigma\bar\sigma \rangle\right)
  \nonumber\\
  |\tau\bar\tau\sigma\rangle &\rightarrow& 
   -(t^2/U) \left(
 |\sigma\tau\bar\tau\rangle + |\sigma\bar\tau\tau \rangle\right)
  \nonumber\\
  |\sigma\tau\bar\tau\rangle &\rightarrow& 
   -(t^2/U) \left(
 |\tau\bar\tau\sigma\rangle + |\bar\tau\tau\sigma \rangle\right),
\end{eqnarray}
where $\sigma$ stands for spins and $\tau$ for empty ($e$) or doubly occupied 
($d$) states, and we use the convention where the creation operators for the 
states are ordered with increasing site index and in case of double occupancy
we use $c^\dagger_{i\uparrow} c^\dagger_{i\downarrow}$.

 In order to calculate the matrix elements describing the transition to
 the upper Hubbard band
 in Eq.~(\ref{eq:nko}), we must 
transform the electron density operator appropriately\cite{oles}, 
$\tilde n_{i,U} = \tilde n^{(1)}_{i,U} + \tilde n^{(2)}_{i,U} + \dots$,
where the leading term 
$  \tilde n^{(1)}_{i,U} = 
    [\tilde T_U, \tilde n_{i,0}]/U$ reads
\begin{eqnarray}
   \tilde n^{(1)}_{i,U} 
    &=& \frac{t}{U} 
    \sum_{\delta=\pm1} 
      \left[
        \tilde n_{i\bar \sigma} 
        \tilde c^{\dagger}_{i \sigma} 
        \tilde c^{\phantom{\dagger}}_{i+\delta \sigma}
      \left(1-\tilde n_{i+\delta,\bar \sigma}\right)
    \right.
    \nonumber\\
    && \left. 
      -  \tilde n_{i+\delta\bar \sigma} 
        \tilde c^{\dagger}_{i+\delta \sigma} 
        \tilde c^{\phantom{\dagger}}_{i \sigma} 
        \left(1-\tilde n_{i,\bar \sigma}\right) 
      \right].
\end{eqnarray}
The next order correction $\tilde n^{(2)}$ will
be considered later.

At half--filling the ground state wave function of ${\cal H}_{\rm eff}$ is
simply that of the Heisenberg model, and matrix elements are with the states
in the upper Hubbard band - i.e. with the states containing exactly one
doubly occupied site. Then 
$\langle f | \tilde n^{(1)}_{j,U} | {\rm GS} \rangle = t/U 
     \langle f | \sum_{\sigma,\delta} 
   (        \tilde c^{\dagger}_{j \sigma} 
            \tilde c^{\phantom{\dagger}}_{j+\delta \sigma} 
          - \tilde c^{\dagger}_{j+\delta \sigma} 
            \tilde c^{\phantom{\dagger}}_{j \sigma} 
    ) | {\rm GS} \rangle $
 holds and ${\cal N}(k,\omega)$ simplifies to
\begin{eqnarray}
  {\cal N}(k,\omega) &=& L \frac{t^2}{U^2} 4 \sin^2 \frac{k}{2} \sum_f 
     \left| \langle f | 
   (        \tilde c^{\dagger}_{1 \sigma} 
            \tilde c^{\phantom{\dagger}}_{0 \sigma} 
          - \tilde c^{\dagger}_{0 \sigma} 
            \tilde c^{\phantom{\dagger}}_{1\sigma} 
    )
 | {\rm GS} \rangle \right|^2 
  \nonumber\\
  && \times
       \delta(\omega - E_f + E_{\rm GS})
       \delta_{k,P_f} .
  \label{eq:nkoeff}
\end{eqnarray}
The density response for this effective model as determined by
exact diagonalization is shown in Fig.~\ref{fig:1} as the dashed
curve.  We can see  that the overall behavior follows that
of the Hubbard model, however there are significant deviations
in the distribution of weights near the edges of the spectrum for  
the not extremely large $U$ used.
This agreement may be improved by including also
the next to leading order correction to the density operator:
\begin{eqnarray}
   \tilde n^{(2)}_{j,U} &=& 
    \frac{1}{U^2} 
     \left[
      \left[\tilde T_U,\tilde T_0\right], \tilde n_{j,0}  \right]
    \nonumber\\
 & = & \frac{t^2}{U^2}\sum_{\sigma,\delta} 
  \bigl[
     (1-2 \tilde n_{j+\delta,\bar\sigma}) 
     (\tilde c^\dagger_{j+2\delta,\sigma}
      \tilde c^{\phantom{\dagger}}_{j,\sigma}
    - \tilde c^\dagger_{j,\sigma}
      \tilde c^{\phantom{\dagger}}_{j+2\delta,\sigma})
      \nonumber\\
    &&+ 2(\tilde c^\dagger_{j+2\delta,\bar\sigma}
       \tilde c^\dagger_{j+\delta,\sigma}
       \tilde c^{\phantom{\dagger}}_{j+\delta,\bar\sigma}
       \tilde c^{\phantom{\dagger}}_{j,\sigma}
       -\tilde c^\dagger_{j,\bar\sigma}
       \tilde c^\dagger_{j+\delta,\sigma}
       \tilde c^{\phantom{\dagger}}_{j+\delta,\bar\sigma}
       \tilde c^{\phantom{\dagger}}_{j+2\delta,\sigma})
  \bigr]
  \nonumber\\
  && + \frac{tV}{U^2} 
       \sum_{\sigma,\delta} 
       \left( 
            \tilde c^{\dagger}_{j \sigma} 
            \tilde c^{\phantom{\dagger}}_{j+\delta \sigma} 
          - \tilde c^{\dagger}_{j+\delta \sigma} 
            \tilde c^{\phantom{\dagger}}_{j \sigma} 
       \right),
\end{eqnarray}
where the final form is correct only when  applied to a half-filled state
with no double occupancy.
The diagonalization result including also these 
terms in the matrix elements is shown as the dotted line  
in Fig.~\ref{fig:1}.  The agreement with the Hubbard model
results as compared to the first approximation 
is noticeably improved as regards the skewing of the
spectrum.

\paragraph{Analytic approach }

The operator 
$\tilde c^{\dagger}_{1 \sigma} 
 \tilde c^{\phantom{\dagger}}_{0 \sigma} 
 - \tilde c^{\dagger}_{0 \sigma} 
 \tilde c^{\phantom{\dagger}}_{1\sigma}$ 
in  Eq.~\ref{eq:nkoeff} removes a spin singlet from the Heisenberg 
wave function and an $ed$ `singlet' is created:
\[
       |\uparrow \downarrow\dots  \rangle - 
       |\uparrow \downarrow\dots  \rangle 
    \rightarrow 2 \left( |de\dots \rangle - |ed\dots \rangle \right).
\]
The wave function of the final state in the large--$U$ limit can be written 
in the product form 
\cite{bogyo}:
\begin{equation}
  | f \rangle = \hat R \left(| \psi \rangle \otimes | \chi \rangle  \otimes 
       | \varphi \rangle\right),
\end{equation}
where $| \psi \rangle$ describes $L-2$ spinless free fermions on an $L$ site
lattice with twisted boundary condition \cite{sorella2penc} (the holes 
represent the $e$ and $d$),
$| \chi \rangle$ is the squeezed wave function of the remaining $L-2$ 
spins with momentum $Q=2 \pi J/(L-2)$, ($J=0,\dots,L-3$) and 
$| \varphi \rangle = (|ed \rangle + |de \rangle)
/ \sqrt{2}$. The operator 
$\hat R = \sum_{j} e^{i \pi n_{j\uparrow} n_{j\downarrow}}$ 
corrects for the opposite sign of hopping of the $d$ compared to $e$.
Using this wave function, for ${\cal N}(\omega,k)$ we get
\begin{eqnarray}
  {\cal N}(k,\omega) &=& L \frac{t^2}{U^2} 16 \sin^2 \frac{k}{2} 
\sum_{Q,I^h_1,I^h_2}
      F_Q B_{01} 
    \nonumber\\
   &&  \times  \delta(\omega - E_f + E_{\rm GS})
       \delta_{k,P_f}.
\end{eqnarray}
The spin part gives 
$F_Q= \sum |\langle \chi^{L-2}_Q| \hat {\cal F} 
  |\chi^{L}_{\rm GS} \rangle |^2$, where the operator  $\hat {\cal F}$ removes
a singlet from the first two sites of the spin wave function and the sum is
over all states with momentum $Q$. This has been studied in
detail in Ref. \onlinecite{talstra}, where 
it is found that $\approx$97\% of the total weight, 
$\sum_Q F_Q = \langle \vec S_0 \cdot \vec S_1 +\case{1}{4}\rangle$ 
($\rightarrow \ln 2$ in the thermodynamic
limit), is concentrated at $Q=0$. This remarkable
feature makes the calculation simple.

  The matrix element coming from the charge part of the wave function is 
$B_{01} = |\langle \psi_f| b_0 b_1| \psi_{\rm GS} \rangle|^2$, where
$b_i$ annihilates a spinless fermion. 
First, we discuss the case $V=0$. Then $|\psi\rangle$
can be characterized by the quantum numbers of the two holes,
$I^h_1$ and $I^h_2$, with momenta $L k^h_{1,2} = 2 \pi I^h_{1,2}+ Q $, and 
the momentum and energy are
\begin{eqnarray}
  P_f &=& -k^h_1-k^h_2+Q = 
          \frac{2 \pi}{L} \left(- I^h_1 - I^h_2 + J\right)\nonumber\\
  E_f &=& E_{\rm GS} + 2 t \left(\cos k^h_1 + \cos k^h_2 \right) +U,
\end{eqnarray}
where we have neglected the small energy difference between the
exchange energies of the ground state and the final states of the order of 
$t^2/U$, furthermore we assume that the ground state has momentum $0$
(i.e. the system has 2,6,10 etc. sites).
Then $B_{01}$ reads
\begin{equation}
  B_{01} = \frac{4}{L^2}
 \sin^2\frac{k^h_1-k^h_2}{2} .
\end{equation}

Let us next consider the case with finite $V$:
The effect of the nearest neighbor repulsion appears as attraction between 
the empty and
double occupied sites , and leads to an extra $-V (1-n_j)(1-n_{j+1})$ term 
in the Hamiltonian describing the spinless fermions.
 Therefore we expect bound
states (excitons) to appear in the spectrum. In the large $U$ limit the 
factorized wave function remains an eigenstate even introducing the 
$V$ term\cite{florian}, supposing that $V \ll U$. 
The two particle problem can be 
solved using the standard Bethe--ansatz 
\begin{equation}
  | \psi \rangle = \sum_{j_1<j_2} a_{j_1,j_2} | j_1,j_2 \rangle ,
\end{equation}
where the holes are located at sites $j_1$ and $j_2$. Since the system is
translationally invariant, we can separate the momentum of the holes $K$
(so that the total momentum is $P_f=-K+Q$):
\begin{equation}
   a_{j_1,j_2} = e^{i (j_1+j_2)K/2} \left[ e^{iq(j_2-j_1)}+\nu
e^{iq(L-j_2+j_1)}\right] .
\end{equation}
 From the twisted boundary condition $a_{0,j}=-e^{-iQ} a_{j,L}$ (we assume that
the number of fermions and holes is even), $\nu =- e^{i (K L/2-Q)} = \pm 1$,
and the quantization $L K=2 \pi I_K + 2Q$ follow, where $I_K$ is an integer.
Then the energy is simply (we introduce for compactness 
$E' = E - U$  and $\omega' = \omega - U$): 
\begin{equation}
 E' =  4 t \cos \frac{K}{2} \cos q , \label{eq:e}
\end{equation}
and the secular equation for $q$ becomes
\begin{equation}
2 \pi I = L q + 2 \arctan \frac{V\cos q +2 t \cos (K/2)}{V \sin q},
\end{equation}
where for $\nu=1$ the quantum number $I$ is integer, $I=0,1,\dots,L/2-1$,  and 
half odd integer ($I=\case{1}{2},\case{3}{2},\dots,(L-3)/2$) for $\nu = -1$. 
Depending on the ratio $V/ [2 t \cos(K/2)]$, we can have either all the 
$q$'s real, or, for sufficiently large ratio a bound state can appear with
complex $q$.
In the thermodynamic limit, $L \rightarrow \infty$, 
the bound state is formed for $V >  2 t \cos \frac{K}{2}$ 
with energy 
\begin{equation}
  E'_{\rm exc} = -V - \frac{4 t^2}{V} \cos^2 \frac{K}{2}.
\end{equation}

$B_{01}$ is now given by $|a_{0,1}|^2/\sum_{j_1<j_2} |a_{j_1,j_2}|^2$ 
and, after a straightforward but tedious calculation, it reads
\begin{equation}
B_{01} 
  = \frac{1}{L} \frac{16 t^2 \cos^2 \frac{K}{2} - E'^2}
     {L(4 t^2 \cos^2 \frac{K}{2} +  V E' + V^2) -  V E' -2 V^2} .
\end{equation}

\begin{figure}
\epsfxsize=8.5 truecm
\centerline{\epsffile{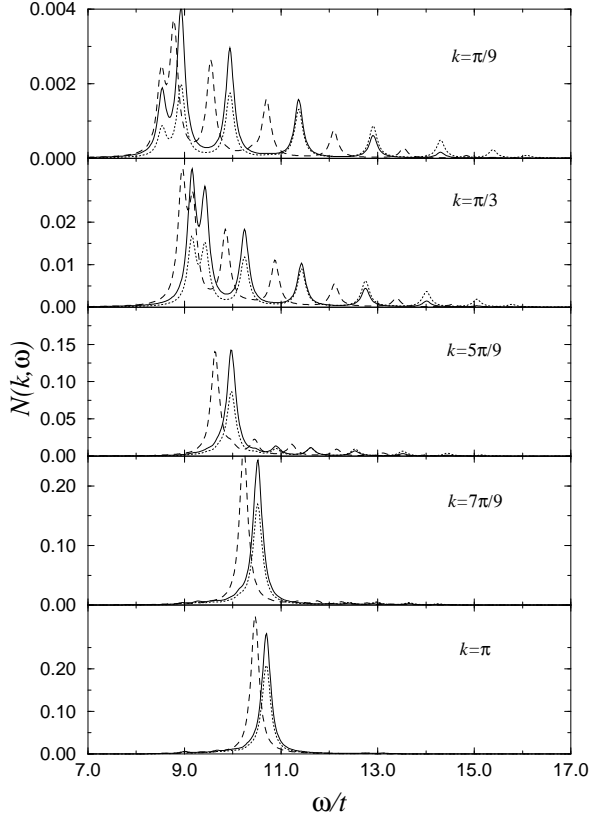}}
\caption{
${\cal N}(k,\omega)$ for an 18 site cluster with $U=12 t$ and $V=2 t$.
The solid line is the ``best'' strong--coupling limit
result, where the exact wave functions and energies from the
diagonalization of ${\cal H}_{\rm eff}$ as well as the $\tilde n^{(2)}$
correction is included, which are however omitted for the dotted line.
The dashed line includes $\tilde n^{(2)}$ but uses the leading
order wave functions and energies for the final states.
}
\label{fig:2}
\end{figure}

  At this point we have the weights to leading order in $t/U$.
We have performed analogous calculations including also 
the term $\tilde n^{(2)}$ in the matrix elements, and the result obtained will
be given later.
Note that there are also contributions of the same order coming
from the wave function correction, which would require going
beyond the factorized wave function.  In order to investigate
the relevance of these terms, we present in Fig.~\ref{fig:2}
a comparison of different approximations.
Apart from an energy shift ${\cal O}(t^2/U)$ the dashed line is
seen to be quite reliable, and is in fact better as far as the
distribution of weights is concerned than is the dotted line,
which shows that the operator corrections are more significant
than those to the wave function.

In the continuum limit we introduce
$g(K,\omega') = L \sum_{I} B_{01} \delta(\omega'-E'_f)$,
and we get 
\begin{eqnarray}
 g(K,\omega') &=& \frac{1}{2\pi}
   \frac{\sqrt{16 t^2 \cos^2 \frac{K}{2} - \omega'^2}}
     {4 t^2 \cos^2 \frac{K}{2} +  V \omega' + V^2} 
     \Theta(4 t \cos\case{K}{2} - |\omega'|) \nonumber\\
  && + \left( 
          1-\frac{4 t^2}{V^2} \cos^2 \frac{K}{2}
       \right)\delta(\omega'-E'_{\rm exc}) ,
\end{eqnarray}
$\int g(k,\omega) d \omega = 1 $ and ${\cal N}(k,\omega)$ now reads:
\begin{equation}
  {\cal N}(k,\omega) \approx \frac{t^2}{U^2} 16 \sin^2 \frac{k}{2} 
      F_0 g(k,\omega-U) + {\cal N}_{\rm inc}(k,\omega) , 
\end{equation}
where ${\cal N}_{\rm inc}(k,\omega)$ contains the contributions from
$Q\ne 0$ and its weight is small (3\%) compared to the main features given by
the $Q=0$ part. 

\begin{figure}
\epsfxsize=8.5 truecm
\centerline{\epsffile{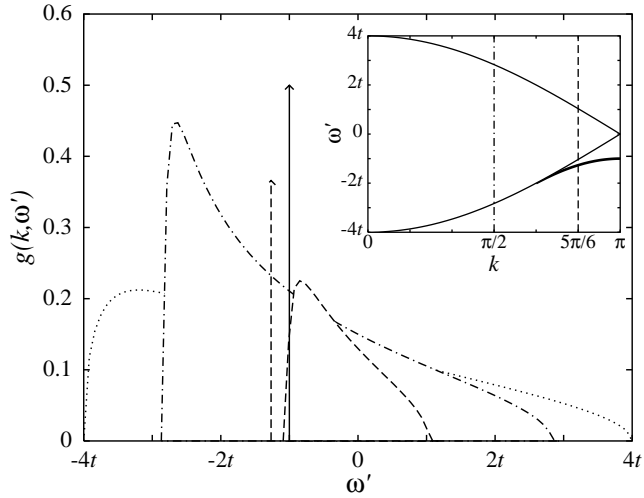}}
\caption{
  $g(k,\omega')$ for $V=t$ and $k=0$ (dotted), $\pi/2$ (dashed--dotted), 
$5 \pi/6$ (dashed) and $\pi$ (solid line). The weight of the excitonic
peak is not in the same scale as the continuum. The inset shows the 
momentum dependence of the boundary the continuum and the exciton dispersion
(heavy line)
}
\label{fig:3}
\end{figure}

The function $g(k,\omega)$ is shown in Fig.~\ref{fig:3}.
As previously alluded to, the spectrum consists  of both
continuum and sharp excitonic features.  For $V \approx t$
the exciton emerges from the continuum in the middle of the
zone, and accounts for almost all of the spectral weight
close to the zone boundary.  For small momentum where the
exciton is ``inside'' the continuum its' presence may still
be seen as a strong enhancement of the intensity of the
spectrum near the lower edge.
Of course for $V > 2t$ the exciton is present for all
momenta.

 The effect of including $\tilde n^{(2)}$ in the operators can be given 
conveniently as 
\begin{equation}
    g(k,\omega) \rightarrow  g(k,\omega) 
   \left(
       1- \frac{4\omega+2V}{U}
   \right)
\end{equation}
for the most important case of $Q=0$.
 For $V=0$, the $t^3U^{-3}$ corrections only 
redistribute the weight and the sum rule
$\int {\cal N}(k,\omega) d \omega = 16 \ln 2 (t/U)^2 \sin^2 k/2 
+ {\cal O}(t^4/U^4)$ is satisfied. 

  To conclude, we have determined the density response of a simple model
 for a correlated quasi--one dimensional insulator. 
The most surprising result is
that excitons may appear near the zone boundary even if they are not
present in the optical absorption (the $k \rightarrow 0$ limit), 
depending on the parameters chosen.
It would be interesting if these features can be observed experimentally.

We acknowledge useful discussions with J. Fink, F. Gebhardt, 
T. Giamarchi and P. Horsch.

\end{document}